\pdfoutput=1

\documentclass[12pt,openright,a4paper]{article}

\usepackage{hyperref}
\usepackage[T1]{fontenc}
\usepackage[utf8]{inputenc}
\usepackage[english]{babel}
\usepackage{amsmath}
\usepackage{amssymb}
\usepackage{caption}
\usepackage{graphicx, floatflt}
\usepackage{booktabs}
\usepackage{textcomp}
\usepackage{subfig}

\usepackage{marginnote}

\begin{document}
\author{Carlo Mengucci,  Daniel Remondini, Gastone Castellani, Enrico Giampieri}

\title{WISDoM: characterizing neurological timeseries with the Wishart distribution} 

\maketitle

\begin{abstract}
WISDoM ({Wishart Distributed Matrices)} is a new framework for the quantification of deviation of symmetric positive-definite matrices associated to experimental samples, like covariance or correlation matrices, from expected ones governed by the Wishart distribution
WISDoM can be applied to tasks of \textit{supervised learning}, like classification, in particular when such matrices are generated by data of different dimensionality (e.g. time series with same number of variables but different time sampling). 
 We show the application of the method in two different scenarios. The first is the ranking of features associated to electro encephalogram (EEG) data with a time series design, providing a theoretically sound approach for this type of studies. The second is the classification of autistic subjects of the ABIDE study, using brain connectivity measurements.
\end{abstract}

\section{Introduction}
High-dimensionality time-structured data are extremely common in fields such as finance, biophysics and biomedical data. 
Very often, experimental limitations lead to uneven sampling (i.e. a different number of time points in terms of frequency or duration) \cite{GCAB} and this poses problems for many types of analysis (e.g. sample classification).
As a consequence, clipping or padding techniques are applied, altering the underlying temporal structure. 
In recent years, studies on such data have seen an increasing popularity in a wide range of fields, from functional magnetic resonance imaging (fMRI) \cite{RNS, SCHZ, GC} to time series  exploration for critical transition prediction in clinical scenarios \cite{SING, SEP}. 
The common goal of this type of research is to develop models and algorithms capable of reaching the highest possible classification and prediction performances, for diagnostic and real time applications, while unveiling underlying information about a system. 
Reproducibility and generalization issues of commonly applied methods are in part caused by \textit{ad-hoc} preprocessing of data, due to the lack of simple null models,  often substituted by reshuffling-based null models.  We introduce a method based on the statistical distribution of symmetric positive-definite matrices (i.e covariance and correlation matrices) extracted from data, using the Wishart distributon as a null model, as a possible way to overcome some of the aforementioned issues. Properties of distribution of random symmetric positive-definite matrices have proven to be useful in fields such as condensed matter, especially in the study of disordered systems \cite{ADA2, CRI_RND}.
The WISDoM  method exploits the properties of the Wishart distribution in order to compute limit distributions for the classes of samples in a classification problem, and a log-likelihood based score is defined for the single variables to quantify their relevance in the classification task.

\section{Method}
\subsection{The Wishart Distribution}
The \textit{Wishart distribution} $ W_p(n, \Sigma)$ is a probability distribution of random nonnegative-definite $p\times p$ matrices that is used to model random covariance matrices.

The parameter $n$ is the number of degrees of freedom (e.g. the number of points in the time series), and $\Sigma$ is a nonnegative-definite symmetric $p\times p$ matrix (with $p$ the number of variables, or \textit{features}, of the time series) called the \textit{scale matrix}.

\textit{Def.} Let $X_1...X_n$ be $N_p(0,\Sigma)$ distribuited vectors, forming a data matrix $p\times n$, $X=[X_1...X_n]$.
The distribution of  a $p\times p$, $M=XX'=\Sigma^n_{i=1}X_iX_i'$ \textit{random matrix}  is a Wishart distribution. \cite{AMS}

We have then by definition:

\begin{equation}
M\sim W_p(n, \Sigma)\sim\Sigma^n_{i=1}X_iX_i' \qquad X_i\sim N_p(0,\Sigma)   
\label{wsim}
\end{equation}

so that $M\sim W_p(n, \Sigma)$  is the distribution of a sum of $n$ rank-one matrices defined by independent normal $X_i \in R^p$ with $E(X)=0$ and $Cov(X)=\Sigma$.

In particular, it holds for the present case:

\begin{equation}
E(M)=nE(X_iX_i') = nCov(X_i) = n\Sigma
\end{equation}

\subsection{PDF Computation for Invertible $\Sigma$}

In general, any $X\sim N(\mu,\Sigma)$ can be represented as
\begin{equation}
X=\mu + AZ,\quad Z\sim N(0,I_p)
\end{equation}
so that
\begin{equation}
\Sigma= Cov(X) = A Cov(Z)A' = AA'
\label{covZ}
\end{equation}

The easiest way to find $ A$ in terms of $\Sigma$ is the LU-decomposition, which finds a unique lower diagonal matrix $A$ with $A_{ii} \geqslant 0 $ such that $AA' = \Sigma$.

Then by \ref{wsim} and \ref{covZ}, with $\mu=0$ we have:
\begin{equation}
W_p(n, \Sigma)\sim \sum_{i=1}^n(AZ_i)(AZ_i)'\sim A(\sum_{i=1}^nZ_iZ_i')A'\sim AW_p(n)A'
\end{equation}
where $Z_i\sim N(0, I_p)$ and $W_p(n)=W_p(I_p,n)$. 

Assuming that  $n\geq p$ and $\Sigma$ is invertible, the density of the random $p\times p $ matrix $M$ in \ref{wsim} can be written 
\begin{equation}
f(M, n, \Sigma)=\frac{1}{2^{\frac{np}{2}}\Gamma_p(\frac{n}{2})\mid \Sigma\mid^{\frac{n}{2}}}\mid M\mid ^{\frac{n-p-1}{2}}exp[-\frac{1}{2}tr(\Sigma^{-1}M)]
\label{Wishart-pdf}
\end{equation} 
\\[10pt]
so that $f(M, n, \Sigma)=0$ unless $M$ is \textit{symmetric and positive-definite}. \cite{IMS}

Note that in \ref{Wishart-pdf} we define $\Gamma_p(\alpha)$ as the \textit{generalized gamma function}:
\begin{equation}
\Gamma_p(\alpha)=\pi^\frac{p(p-1)}{4}\prod_{i=1}^p\Gamma(\frac{2\alpha+1-i}{2})
\end{equation}

\subsection{Estimation of the Wishart Parameters from Empirical Covariance}

We justify the use of the Wishart distribution  under the assumption of Multivariate Gaussian distributed data scenarios.
This kind of assumption is indeed generally good for a wide range of problems.
Furthermore, the use of the average covariance matrix (obtained from all the elements of one class) to compute the scale matrix for the class estimated distribution will be proven to be a good approximation of a complete Bayesian model.

This is done by considering that the Wishart Distribution is the conjugate prior of a multivariate Gaussian distribution, such as the Gamma distribution for the univariate Gaussian case.
By considering a Gaussian model with known mean $\mu$, so that the free parameter is the variance  $\sigma^2$, as in \cite{SML}, the likelihood function is defined as follows:
\begin{eqnarray}
p(X_1...X_n\mid \sigma^2)\propto (\sigma^2)^{-\frac{n}{2}}exp(-\frac{1}{2\sigma^2}n \overline{(X-\mu^2)}) , \\ \overline{(X-\mu^2)}=\frac{1}{n}\sum_{i=1}^n(X_i-\mu)^2
\label{gausslike}
\end{eqnarray}
\\[12pt]
The conjugate prior is an inverse Gamma distribution. Recall that $\theta$ has an inverse Gamma distribution with parameters $(\alpha,\beta)$   when $\frac{1}{\theta}\sim Gamma(\alpha,\beta)$.

The density  then   takes the form 
\begin{equation}
\pi_{\alpha,\beta}(\theta)\propto \theta^{-(\alpha+1)}e^{-\frac{\beta}{\theta}}
\end{equation}

Using this prior, the posterior distribution of $\sigma^2$ is given by
\begin{equation}
p( \sigma^2 \mid X_1...X_n)\sim InvGamma(\alpha + \frac{n}{2}, \beta + \frac{n}{2} \overline{(X-\mu^2)})
\end{equation}

In the multidimensional setting, the inverse Wishart takes the place of the inverse Gamma.
It has already been stated that the Wishart distribution is a distribution over \textit{symmetric positive semi-definite} $d\times d$ matrices $W$. A more compact form of the density is given by
\begin{eqnarray}
\pi_{\nu_0, S_0}(W)\propto \mid W\mid ^{\frac{(\nu_0-d-1)}{2}}exp(-\frac{1}{2}trace(S_0^{-1}W)), \\ \mid W\mid=det(W)
\end{eqnarray}

where the parameters are the degrees of freedom $\nu_0$ and the positive-definite \textit{scale matrix}
$S_0$.

If $W^{-1}\sim Wishart(\nu_0, S_0)$ we can then state that $W$ has an \textit{Inverse Wishart Distribution}, whose density has the form
\begin{equation}
\pi_{\nu_0, S_0}(W)\propto \mid W\mid ^{-\frac{(\nu_0+d+1)}{2}}exp(-\frac{1}{2}trace(S_0W^{-1})), 
\end{equation}
Let $X_1...X_n$ be $N(0,\Sigma)$ distributed observed data. Then an inverse Wishart prior multiplying the likelihood $p(X_1...X_n\mid \Sigma)$ yields

\begin{align}
& p(X_1...X_n\mid \Sigma)\pi_{\nu_0, S_0}(\Sigma)\propto \\ 
&\mid\Sigma \mid ^{-\frac{n}{2}}exp(-\frac{n}{2}tr(\overline{S} \Sigma^{-1}) \mid \Sigma\mid ^{-\frac{(\nu_0+d+1)}{2}}exp(-\frac{1}{2}tr(S_0\Sigma^{-1})) \\ 
&= \mid \Sigma\mid ^{-\frac{(\nu_0+d+n+1)}{2}}exp(-\frac{1}{2}tr((n\overline{S}+S_0)\Sigma^{-1}))
\end{align}
\\[10pt]
where $\overline{S}$ is the \textit{empirical covariance} :
\begin{equation}
\overline{S}=\frac{1}{n}\sum_{i=1}^nX_iX_i^T
\end{equation}

Thus, an a posteriori distribution with the form
\begin{equation}
p(\Sigma\mid X_1...X_n)\sim InvWishart(\nu_0+n, n\overline{S}+S_0)
\end{equation}

is obtained.

Similarly, it can be stated that for the inverse covariance (\textit{precision}) matrix $\Sigma^{-1}$ the conjugate prior is a Wishart distribution.

\subsection{Class-Wise Estimated Distribution}

The core idea of the WISDoM method is to represent each element undergoing classification as a covariance matrix of its features. 
Nominally, each element can be characterized by the covariance matrix extracted by the repeated observations of the vector of its features, for example derived by a time series.
The aim is to use the \textit{free parameters} of the Wishart distribution (the scale matrix $S_0$ and the number $n$ of the degrees of freedom, as shown in \ref{Wishart-pdf}) to compute an estimation of the distribution for a certain class of elements, and then assign a single element to a given class by computing a log-likelihood between the element being analyzed and each class.
Furhermore,
a score can be assigned to each feature by estimating the 
variation in terms of log-likelihood, due to its removal from the feature set. 
If the removal of a feature causes significant increase (or decrease) in the log-likelihood, it can be stated that such feature is highly representative of the system analyzed.
Thus, the WISDoM approach allows  not only to assign a given element to a class, but also to identify the features with the highest relevance in the classification process.

Covariance matrices are a good choice for a distance metrics in a classification task, both for the way they represent a system and for the property that the average of a set of covariance matrices is a covariance matrix itself. 
If each element of a given class $C$ is represented by a covariance matrix $\Sigma$ of its features, this property allows us to estimate a distribution for the class by choosing 
\begin{equation}
S_0=\hat{\Sigma}_C=\frac{1}{N}\sum_{i=1}^N \Sigma_i
\label{Class-Scale}
\end{equation}

The other necessary parameter for the estimation is the \textit{number of degrees of freedom n}.
Assume that an $X_i=(x_1, ..., x_p)$ vector of $p$ features is associated to each element $i$ of a given class, while having $n$ observations for this vector. 
The covariance matrix $\Sigma_i$ computed over the $n$ observations will represent the interactions between the features of element $i$.
The number of degrees of freedom  $n$ of the Wishart distribution is then given by the number of times $X_i$ is observed.

Let us give an example tied to \textit{functional MR brain imaging}. 
An image of patient $i$'s brain is acquired; as usual these images are divided in a certain number $p$ of zones (voxel, pixel etc.), each zone being sampled $n$ times over a given time interval in order to observe a certain type of brain activity and functionality.
In this example, the features contained in vector $X_i=(x_1,..,x_p)$ associated to patient $i$ are indeed the zones chosen to divide the brain image, each zone having been sampled $n$ times during an acquisition interval.
The  $p \times p$ correlation matrix $\Sigma_i$  is then representative of the functional correlation between the $p$  brain areas.
Repeating this procedure for the $N$ patients of a known class $C$ (i.e. a diagnostic group) and computing the $\hat{\Sigma}_C$ scale matrix for the class, will allow us to estimate a Wishart distribution for that class  and draw samples from it.

\subsection{Log-Likelihood Ratio Score}

After defining how to represent classes distribution, WISDoM allows to compute the log-likelihood of each element to belong to one of the classes.
Moreover, WISDoM allows to compute the variation of log-likelihood ratio scores due to the removal of features, singularly or in groups, thus estimating how much the classification performance changes. Uninformative (or less informative) features can thus be pruned, allowing for a dimensionality reduction of the initial feature set.
The whole process can be seen as a feature transformation, mapping the covariance matrix $\Sigma_i$ of subject $i$ to a score vector formed by the change in log-likelihood for each feature.

\subsubsection*{Complete Matrix Score}

The  WISDoM Classifier relies upon computing the log-likelihood of a matrix $\Sigma_i$ with respect to the Wishart distribution estimated for a class $C$, using $\hat{\Sigma}_C$ as the scale marix.
If a problem concerning two given classes $C_A$ and $C_B$ is taken into account, the score assigned to each $\Sigma_i$  can be defined as follows:

\begin{equation}
score_{i}= logP_{W}(\Sigma_{i}\mid n, \hat{\Sigma}_{A})-logP_{W}(\Sigma_{i}\mid n,\hat{\Sigma}_{B})
\label{pdf-score}
\end{equation}

Where $\hat{\Sigma}_{A,B}$ are the scale matrices computed for the classes $A,B$ respectively, and $ logP_{W}(\Sigma_{i}\mid n, \hat{\Sigma}_{A,B})$ is the logarithm of the probability of $\Sigma_i$ belonging to the Wishart distribution estimated for one of the two classes $A,B$.

\subsubsection*{Single Feature Score}

WISDoM allows to obtain information about the features used for classification by reducing the matrix $A$ to its \textit{principal submatrices} (see  Appendix).
An important property for the principal submatrices of a symmetric positive definite matrix is that \textit{any $(n-k)\times (n-k)$ partition is also symmetric and positive definite.}

By removing one feature from the dataset, calculating the WISDoM scores, and iterating this process over all the features (i.e. analyzing all  the $(p-1)\times (p-1)$ principal submatrices of $\Sigma_i$ and $\hat{\Sigma}_C$) the method can assign a score to each feature, representing its relevance in the decision for $\Sigma_{i}$ to be assigned to one class or another.
Let $\Sigma_j$ be a principal submatrix of order $(p-1)$, of the matrix $\Sigma$ computed on the observation of $X_i=(x_1,...,x_p)$ for subject $i$, \textit{obtained by the deletion of the $j^{th}$ row and the $j^{th}$ column}. 
\vspace{5mm}
Similarly, let  $\hat{\Sigma}_{Cj}$  be a principal submatrix of order $(p-1)$, of the matrix $\hat{\Sigma}_C$ computed for the class $C$. 
The score assigned to each feature of $X_i=(x_1,...,x_p)$ is then given by eq.(\ref{JSscore}).
\begin{align}
&Score_j(C)=\Delta logP_{Wj}(C)= \\
&logP_W(\Sigma, n \mid \hat{\Sigma}_C,n)-logP_W(\Sigma_j, n \mid \hat{\Sigma}_{Cj},n)
\label{JSscore}
\end{align}
\vspace{5mm}
In a 2-class example, we obtain a score vector as follows:
\begin{equation}
Ratio_j=\Delta logP_{Wj}(C_1)-\Delta logP_{Wj}(C_2)
\label{JRatio}
\end{equation}

A generalization to $(p-n)$ dimensionality reduction can be found in the supplemental materials.

\section{Results}

\subsection{Eye state detection via EEG}
\begin{figure}[!h]
\includegraphics[scale=0.45]{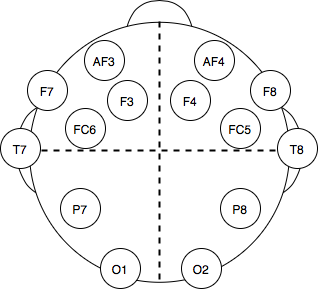}
\caption{\textit{Electrodes position in the headset used for EEG dataset acquisition.}}
\label{headset}
\end{figure}

The dataset used was downloaded from the UCI Machine Learning Repository (\textbf{http://archive.ics.uci.edu/ml/}).
This dataset has been chosen for many reasons: it's openly accessible, contains records from 14 electrodes with standard headset placement (fig.\ref{headset}),  thus making the features of our problem directly linked to brain topology and a published classification performance benchmark on the dataset exists \cite{UCI}.
The data consisted in a series of 14980 time points, sampled for each one of the 14 electrodes and labelled with a 1 or a 0 to mark wether the eyes of the subject are open or closed at that time point. 
The time series has been split into batches of different length according  to eye state changes. In this way, a correlation matrix can be extracted for each batch (the "elements" for this classification problem), while the length of each batch is used for computing the degrees of freedom of each class Wishart distribution during training.
A total of 140 batches with various lenghts, 70 with eye state 1 and 70 with eye state 0, were obtained.

The representative matrix for each class is computed as the  average (weighted on the length of each batch) of matrices of the elements belonging to eye state 0 or eye state 1, excluding the element to be predicted in a \textit{Leave One Out} fashion in order to avoid overfitting. 
By doing this, we verify that the method is independent from the sampling window chosen when applied to time series data, with the only constraint that the length of such window cannot be less than the number of the features of the system.

After undergoing feature score computation, a stochastic grid search on a set of classifiers has been performed in order to obtain the best prediction performance with the transformed features. All the classification tasks are validated through a \textit{10-fold crossvalidation}.
Versions and references for all Python packages used can be found in supplemental and at \cite{scipy, pandas, matplotlib, numpy, sympy, jupyter, ipython, seaborn}.

\begin{figure}[!h]
\includegraphics[scale=0.4]{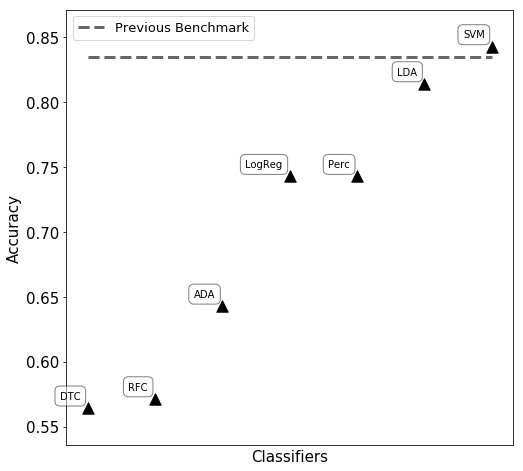}
\caption{\textit{Performance comparison of different classifiers on the WISDoM EEG transformed features. The classifiers are reported as follows: RFC: Random Forest Classifier \cite{RFC}, DTC: Decision Tree Classifier\cite{DTC,DTC2}, ADA: ADA Boosting Tree Classifier \cite{ADA, ADA2}, LDA: Linear Discriminant Analysis Classification \cite{LDA}, LogReg: Logistic Regression Classifier \cite{logreg}, Perc: Perceptron Classifier \cite{perc}, SVM: C-Support Vector Machine\cite{SVM, SVM2}. All classifiers are SciKitLearn implementations. } }
\label{class_com}
\end{figure}

We first tried to assess eye state using complete matrix score, as in eq. (\ref{pdf-score}). Classifiers reported in fig.(\ref{class_com}) were trained and tuned, with the aim of obtaining the best performance possible. However, in this scenario the resulting classification performances were poor, reaching an accuracy of $\sim60\%$ in the best cases. We then proceeded to compute single feature scoring, as in eq.(\ref{JSscore}), obtaining a feature transformation.
As in fig.(\ref{class_com}), different classifiers belonging  to two main categories (decision trees and linear classifiers) have been trained on the transformed features. 
The best performance has been achieved with a C-support Vector Machine (\textit{Python 3.6 SciKitLearn implementation)} resulting in  a $0.85\%$ ROC AUC score and an accuracy score of $84.3\%$, comparable with
the benchmark of $83.5 \%$ accuracy set by Rajesh Kanna et al.\cite{UCI}.

\begin{figure}[!h]
\includegraphics[scale=0.35]{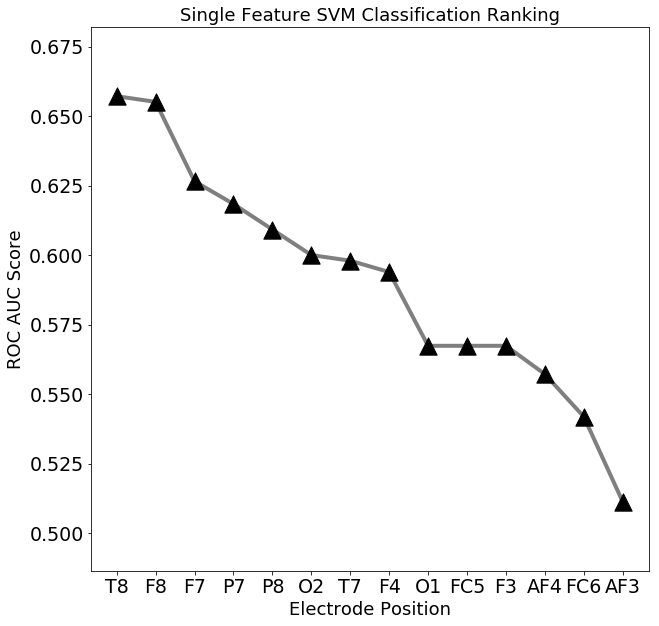}
\caption{\textit{EEG feature ranking based  classification performance (ROC AUC score). Temporal (T8) and outer frontal (F8, F7) electrodes seems to convey the most important signals for eye state prediction. }}
\label{EEG-perf}
\end{figure}

\begin{figure}[!h]
\includegraphics[scale=0.4]{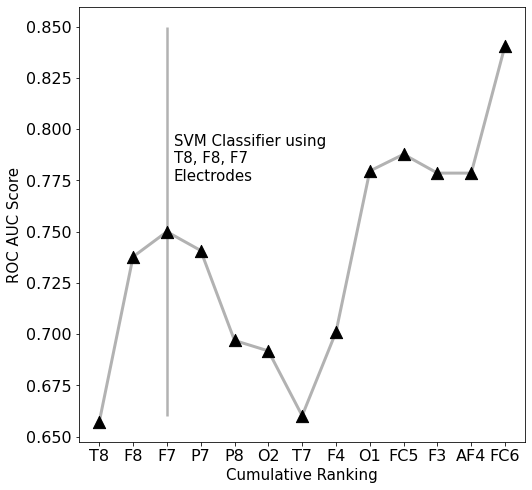}
\caption{\textit{EEG features performance landscape at an increasing number of ranked features used for classification. Labels on the X-axis point out which feature is being added to the previous ones.}}
\label{ROC_N_rank}
\end{figure}

To assess which features contain the largest amount of useful information for prediction, a set of single feature C-SVM classifications has been performed (fig.\ref{EEG-perf}): a performance of $~75\%$ accuracy is obtained  by using only the top three ranking electrodes (fig. \ref{ROC_N_rank}). Training the classifier with the top three electrodes yields a local maxima in the landascape performance, highlighting the importance of the information recorded by these three electrodes about the state of the whole system.

\subsection{Autism classification via fMRI}

The Autism Brain Imaging Data Exchange \cite{abide} is a consortium effort, aggregating fMRI  datasets from individuals with autism spectrum disorders (ASD) and age-matched typical controls (TC). 
Data from 17 acquisition sites were merged, using different preprocessing tools and pipelines \cite{abidepreproc}. Complete information about the dataset is found at \textit{http://preprocessed-connectomes-project.org/abide/}. 
For our classification task , we focused on male subjects of the "Autism" diagnostic group (AUT): we analyzed a total of 369 TC and 220 AUT subjects, with 200 time points each (number of degrees of freedom of the Wishart distribution). The chosen preprocessing pipeline for the extraction of the average time series of the ROIs is the  Connectome Computation System (CCS), with a global signal correction and the application of a bandpass filter (0,01-0,1 Hz). The 116 ROIs (features), of which covariance matrices are extracted, are labelled according to the Automated Anatomical Labeling of the Brain (AAL2) \cite{aal2}. 

The representative matrix for each class is computed as the  average of matrices of the elements belonging to class AUT or TC, excluding the sets of element to be predicted. For this task, a shuffled 10-fold splitting of the dataset for feature score computation has been used to avoid overfitting. 
After undergoing feature score computation, the parameters of a C-Support Vector Machine Classifier have been fine tuned in order to obtain the best prediction performance with the transformed features. All the classification tasks are validated through a \textit{stratified 10-fold crossvalidation}, in order to minimize the effects of class imbalance in train and test sets. 
Versions and references for all Python packages used can be found in supplemental and at \cite{scipy, pandas, matplotlib, numpy, sympy, jupyter, ipython, seaborn}.

\begin{figure}[!h]
\includegraphics[scale=0.4]{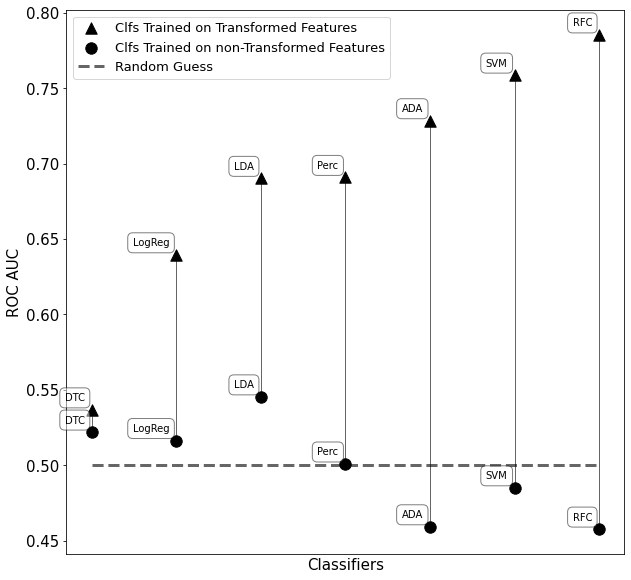}
\caption{\textit{Performance comparison of different classifiers on the WISDoM transformed features and non-transformed feature in the ABIDE dataset. The classifiers are reported as follows: RFC: Random Forest Classifier \cite{RFC}, DTC: Decision Tree Classifier\cite{DTC,DTC2}, ADA: ADA Boosting Tree Classifier \cite{ADA, ADA2}, LDA: Linear Discriminant Analysis Classification \cite{LDA}, LogReg: Logistic Regression Classifier \cite{logreg}, Perc: Perceptron Classifier \cite{perc}, SVM: C-Support Vector Machine\cite{SVM, SVM2}. All classifiers are SciKitLearn implementations. } }
\label{ABIDE_class_com}
\end{figure}

The C-SVM classifier  trained on the transformed features resulted in an accuracy score of $72.1\%$ and a ROC AUC score of $0.76$. Furhtermore, we obtained a ROC AUC score of $0.79$ and an accuracy of $73.5\%$ with a fine tuned Random Forest Classifier. We also compared the classifiers in fig.\ref{class_com} training them with WISDoM transformed features and non-transformed features (nominally, the elements of the lower triangle of each covariance matrix). Results in fig.\ref{ABIDE_class_com} show an overall improvement of classification performances when using transformed features.
As a comparison, the state of art of classification on the entire ABIDE dataset is set at $70\%$ accuracy obtained with a deep learning architecture built by Heinsfeld et al. \cite{abidednn}. This result on the whole spectrum of autism required the use of various stacked denoising autoencoders and hidden layers, resulting in a large time-consuming training routine ($\sim 33$ hours), while WISDoM obtained satisfying classification performance in much smaller time ($\sim18$ minutes, including feature transformation which is the most time-consuming step of the pipeline). 

\section{Discussion}
The WISDoM framework is introduced: a method for modelling symmetric positive definite matrices, such as covariance  and correlation matrices, used in a wide array of problems. 
It can provide a null model for classification purposes in which each sample is represented as a  covariance/correlation matrix, even if the number of observations (e.g. the length of the time series) is different from sample to sample.
This property makes the WISDoM method suitable for problems with non-homogeneous data size, for example time series with uneven lengths, missing points or irregularly sampled data.
Moreover, we show that a feature transformation based on WISDoM scores can be used for dimensionality reduction, providing a ranking for the most important variables in the dataset.
While showing good generalization capabilities with time-series data and non-homogeneous sampling related issues, the method is not suitable when the number of features exceeds the sampling ($p>n$). This is a theoretical limit tied to the invertibility of the scale matrix required to compute the Wishart probability density function. At present, WISDoM cannot thus be applied to problem involving the so called "long data", such as gene expression tables, unless considering corrections such as matrix regularization methods and hierchical methods such as power priors \cite{powerp}

The method has been tested on the EEG eye state prediction dataset of the open \textit{UCI Machine Learning Repository}, slightly improving the previous classification benchmark with little to no preprocessing, and giving useful insights on the minimum number and location of electrodes needed to record sufficient information for the task.
Moreover, the method has been applied to the classification of a subset of the ABIDE dataset, using brain functional connectivity data. We obtained satisfying classification scores, comparable with the state of art classification results on the dataset, with very simple classifiers and without the use of additional time-consuming processing routines. Furthermore, the Bayesian-like framework of scores-computation through log-likelihood, could allow for a sort of inline learning by continuously updating the estimation of each class Wishart distribution. This property makes the WISDoM method also suitable for real-time learning during data acquisition.

\section*{Author contribution}
EG, CM, GC and DR designed the research. CM and EG analyzed the data and implemented the method. CM, DR, GC and EG wrote the paper.

\bibliographystyle{ieeetr}
\bibliography{WISDoM-arxiv-update} 

\section*{Figure captions}

\end{document}